\documentclass[%
 reprint,
 amsmath,amssymb,
 aps,
]{revtex4-2}
\usepackage[utf8]{inputenc}	% Para caracteres en español
\usepackage[skip=2pt,font=scriptsize]{caption}
\usepackage{graphicx}% Include figure files
\usepackage{dcolumn}% Align table columns ondecimal point
\usepackage{subcaption}
\usepackage{caption}
\usepackage{amsthm}
\usepackage{bm}% bold math
\usepackage{verbatim} 
\usepackage{xcolor}
\usepackage{lipsum}
\usepackage{comment}

\begin{document}
\title{Renormalization Group-Motivated Learning}

\author{Jonathan Landy}
\email{jslandy@gmail.com}
\affiliation{Stitch Fix, Inc.}

\author{Tsvi Tlusty}
\email{tsvitlusty@gmail.com}
\affiliation{Center for Soft and Living Matter, Institute for Basic Science (IBS), Ulsan 44919, Korea}
\affiliation{Department of 	Physics, Ulsan National Institute of Science and Technology (UNIST), Ulsan 44919, Korea}

\author{YeongKyu Lee}
\affiliation{Department of Physics, Gyeongsang National University}

\author{YongSeok Jho}
\email{ysjho@gnu.ac.kr}
\affiliation{Department of Physics, Gyeongsang National University}

\date{\today}

\begin{abstract}
We introduce an RG-inspired coarse-graining for extracting the collective features of data. The key to successful coarse-graining lies in finding appropriate pairs of data sets. We coarse-grain the two closest data in a regular real-space RG in a lattice while considers the overall information loss in momentum-space RG. 
Here we compromise the two measures for the non-spatial data set. 
For weakly correlated data close to Gaussian, we use the correlation of data as a metric for the proximity of data points, but minimize an overall projection error for optimal coarse-graining steps. 
It compresses the data to maximize the correlation between the two data points to be compressed while minimizing the correlation between the paired data and other data points. 
We show that this approach can effectively reduce the dimensionality of the data while preserving the essential features.
We extend our method to incorporate non-linear features by replacing correlation measures with mutual information. This results in an information-bottleneck-like trade-off: maximally compress the data while preserving the information among the compressed data and the rest. 
Indeed, our approach can be interpreted as an exact form of information-bottleneck-like trade off near linear data. 
We examine our method with random Gaussian data and the Ising model to demonstrate its validity and apply glass systems. Our approach has potential applications in various fields, including machine learning and statistical physics.
\end{abstract}
\maketitle

\section{Introduction}
Phase transition and critical phenomena are among the greatest achievements in statistical mechanics in the last century.
The concepts have been extensively incorporated in various fields, including biology~\cite{munoz2018colloquium}, economics~\cite{castellano2009statistical}, and even deep learning~\cite{tishby2015deep}.  
While the physical concepts get more attraction across different areas because they are not only conceptually captivating and but also practically useful, one of the most important theoretical achievements for studying the phase transition and critical phenomena, renormalization group (RG) theory~\cite{wilson1983renormalization,li2018neural,kardar2007statistical,parisi1988statistical}, has been limited in application.

Recently, there were successes in connecting RG and learning.
Mehta and Schwab are one of the first to introduce RG to uncover the underlying mechanisms of deep learning~\cite{mehta2014exact}. Li and Wang utilize RG in the learning model of deep neural networks~\cite{li2018neural}. 
Bialek showed the relevance between the PCA and RG~\cite{bradde2017pca}, and the RG-like approach could be applied to the large-scale correlated behavior of time-varying neuro-signals ~\cite{meshulam2019coarse,meshulam2017collective}. 
These works show that RG can play a key role in extending phase transition picture to various fields. 
In this work, we will utilize the RG picture to find collective behaviors of complex data. Especially we focus on the time-series data, such as molecular dynamics trajectories.

Molecular dynamics is a powerful tool for monitoring the dynamics of complex systems at a molecular level~\cite{frenkel1996understanding}, but, interpreting simulation results becomes increasingly challenging as systems become more and more complex.
For instance, the collective properties from the glass simulation results are challenging to diagnose~\cite{keys2011excitations,schoenholz2016structural} despite the well-established simulation protocol for making glass~\cite{kob1995testing}. 
Though several macroscopic parameters are defined to capture collective behaviors, including machine-learning aid parameters~\cite{widmer2006free,widmer2008irreversible,lee2020local}, they are inconclusive.

The molecular dynamics simulation generates time-series trajectories of particle positions, from which we may extract the collective features of the particles. 
However, because the positions of the particles are continually changing, application of the real space RG approach may not be feasible. 
Momentum space RG may be an alternative, but it has the drawback of losing spatial information. 
As a compromise, we take advantage of both approaches through a coarse-graining step that averages two data similar to real space RG but in a way minimizing the total loss of information quantity. We describe a fast step-wise procedure for carrying out linear analysis coarse-graining for a general data set, and provide an open source implementation of this at \cite{code_location}.

This paper is organized as follows: we start with a nearly Gaussian data set. For these weakly correlated data, the linear analysis works fine. Then, we extend the linear coarse-graining rule to a non-linear one, which eventually be expressed as an information bottleneck theory. Finally, we provide numerical examples to demonstrate the validity of our approach. 

\section{Coarse Graining}
In the RG process, we may need to determine two rules for each step: 1) how to reduce the dimension of the data by summing two data with proper scaling, and 2) how to select coarse-graining pairs. However, in our purpose, the exact scaling may not be critical as long as we can track down the scale changes. 

Suppose we have a set of $2N$ variables $\{x_i\}$ for $i=1,2,\cdots,2N$. Then, the partition function of the data set is,
\begin{equation}
    Z \equiv \text{Tr}_{x_i} e^{-H\left(\left\{x_i\right\}\right)}.
\end{equation}

The system, approximated using $N$ coarse-grained variables $\{z_j\}$ for $j=1, 2, \cdots, N$, satisfies, 
\begin{eqnarray}
    e^{-H_{\omega}^{RG}\left(\left\{ z_j\right\}\right)} \equiv \text{Tr}_{\{x_i\}} e^{T_{\left\{\omega\right\}} \left( \left\{x_i\right\},\left\{z_j\right\}\right) - H\left(\left\{ x_i\right\} \right)}.
\end{eqnarray}

If we only consider the fluctuations near Gaussian, the probability of $\{x_i\} ,\left\{z_j\right\}$ is proportional to,

\begin{widetext}
  \begin{eqnarray}
    e^{-\frac{1}{2}\left[ K_{11}x_1^2+ K_{12}x_1x_2 + \cdots + K_{(2n)(2n)}x^2_{2n} +\left(x_1 + x_{n+1} \right) \omega_{1(n+1)}z_1  + \cdots + \left(x_n + x_{2n}\right) \omega_{(n)(2n)}z_n \right] }. 
  \end{eqnarray}
\end{widetext}

We assume that the coarse-graining step summates two random variables with proper scaling $\omega$,
\begin{widetext}
  \begin{eqnarray}
    e^{-H_{\omega}^{RG}\left\{ z_j\right\}} &=& \int \prod_i^{2n} dx_i  e^{-\frac{1}{2}\sum_{ij}  x_i^T K_{ij} x_j} \prod_{j=1}^{n} \delta\left(z_j - \omega(x_i+x_{i+n})\right)\nonumber \\
    &=& \frac{1}{(2\pi)^{n}}\int \prod_i^{2n} dx_i \int \prod_{l=1}^{n} dk_l
    e^{-\frac{1}{2}\sum_{ij}  x_i^T K_{ij} x_j} e^{-i k_l \left(z_j - \omega(x_i+x_{i+n})\right)} \nonumber \\
    &=& \frac{1}{\sqrt{\det K}}  \int \prod_{l=1}^{n} dk_l
    e^{-\frac{\omega^2}{2}\sum_{i,j}^{n}  k_i^T \left( K^{-1}_{ij} + K^{-1}_{(i+n)j} + K^{-1}_{i(j+n)} + K^{-1}_{(i+n)(j+n)}\right) k_j} e^{-i k_l z_j}
  \end{eqnarray}
\end{widetext}

If we set $\left(K^{RG}\right)_{ij}^{-1} = K_{ij}^{-1} + K_{(i+n)j}^{-1} + K_{i(j+n)}^{-1} + K_{(i+n)(j+n)}^{-1} $,
\begin{widetext}
  \begin{eqnarray}
    e^{-H_{\omega}^{RG}\left\{ z_j\right\}} 
    &=& \frac{1}{\sqrt{\det K}}\int \prod_{l=1}^{n} dk_l
    e^{-\frac{\omega^2}{2}\sum_{i,j}^{n}  k_i^T \left(K^{RG}\right)_{ij}^{-1} k_j} e^{-i k_l z_j} \nonumber \\
    &=&  \left(\frac{\det K^{RG}}{\det K}\right)^{\frac{1}{2}} \left(\frac{\sqrt{2\pi}}{\omega}\right)^{n}
    e^{-\frac{1}{2\omega^2}\sum_{i,j}^{n}  z_i^T \left(K^{RG}\right)_{ij} z_j}.
  \end{eqnarray}
\end{widetext}

If we normalize the mean of ${x_i}$ to be zero, so is the mean of ${z_j}$.
Assumption that the variance remains the same during the process yields,
\begin{equation}
    \omega^2 \left(K^{RG}\right)_{ii}^{-1} = K_{ii}^{-1}.
\end{equation}

For a very weakly correlated system, we may assume that $K_{ii}^{-1} \gg K_{ij}^{-1} $ for $i\neq j$. 
\begin{eqnarray}
    2\omega^2  \simeq 1 \text{ or }
    \omega \simeq \frac{1}{\sqrt{2}}
\end{eqnarray}
This corresponds to the central limit theorem.
In this way, we rescale the data setting the mean of the coarse-grained data zero and the variance constant.

\underline{Mutual Information}
Mutual information is a quantity to evaluate the loss of information during transmission for nonlinear data.
Minimizing the mutual information change during the coarse-graining step,
\begin{eqnarray}
    I_{\omega}\left({z_j}\lvert{x_i}\right) = Tr_{z_j,x_i} P_{\omega}\left({z_j},{x_i}\right)\log\left(\frac{P_{\omega}\left({z_j},{x_i}\right)}{P_{\omega}\left({z_j}\right)P\left({x_i}\right)}\right),
\end{eqnarray}
may be a way to find the optimal coarse-graining step.
But, the mutual information diverges due to the delta function term in the logarithmic function. 
There are suggestions that avoid the divergence by keeping the essential property of the mutual information. For example, Koch-Janusz considers mutual information between the coarse-grained variables and the environmental variables which are not involved in the coarse-graining~\cite{koch2018mutual}. 

It may be necessary to redefine the information quantity for the coarse-graining process to account for the non-linear coupling of the data~\cite{sarra2021renormalized}, as we will describe in detail later in the paper.

\subsection{Making a Set of Coarse-Graining Pairs for Linear Analysis}
In the linear analysis, the information quantities correspond to the correlations. 
We express the coarse-graining error using the correlations. 

The coarse-graining step involves the pair selection rule. In real space RG, we typically coarse-grain two nearest neighbors in order to retain spatial correlation information. However, we are dealing with data where spatial information has been discarded or ambiguous. Theoretically, there are a vast number of ways to choose pairs roughly on the order of ${N \choose 2}{N-2 \choose 2}\cdots{4 \choose 2}{2 \choose 2} = \frac{N!}{2^N} \simeq \left(\frac{N}{2}\right)^N$, which is computationally impractical for large $N$.

One option to tackle this issue is establishing proximity measures between data points. However, for our purposes it may be more appropriate to use divergence metrics between probability distributions.

We start with a linear measurement of the coarse-graining error.

\underline{Projection to $N$ dimension}
Recently, Serena Bradde, William Bialek analyzed PCA in the context of momentum space RG~\cite{bradde2017pca}. They demonstrated that the coarse-graining error corresponds to the error in projecting data onto the principal components. 

PCA uses a linear transformation to project $N_0$-dimensional correlated data to $N_1$-uncorrelated dimensions, when $N_1 \le N_0$.
While it provides an optimized projection among linear transformations, interpreting the projected axes is difficult since they are linear combinations of the original data axes.
A more straightforward approach is to use the original axes as the projection axes~\cite{landy2017stepwise}. 
The first step is to project the matrix $X$($\in \mathrm{R}^{2n\times m}$ where $m$ data, $2n$ features) onto half-sized matrix $Y$($\in \mathrm{R}^{n\times m}$).

For original matrix $X \in R^{2n\times m}$ where, $2n$ features and $m$ data, and projected matrix $Y \in R^{n\times m}$,
\begin{widetext}
  \begin{eqnarray}
    X &=& 
    \begin{bmatrix}
        x_{11} & x_{12}  & \cdots & x_{1(m)} \\ 
        x_{21} & x_{22}  & \cdots & x_{2(m)} \\ 
        \vdots & \vdots  & \cdots & \vdots \\ 
        x_{(2n)1} & x_{(2n)2}  & \cdots & x_{(2n)(m)} \\
    \end{bmatrix} 
    X^T =
    \begin{bmatrix}
        x_{11} & x_{21}  & \cdots & x_{(2n)1} \\ 
        x_{12} & x_{22}  & \cdots & x_{(2n)2} \\ 
        \vdots & \vdots  & \cdots & \vdots \\ 
        x_{1m} & x_{2m}  & \cdots & x_{(m)(2n)} \\
    \end{bmatrix}\\
    Y &=& \frac{1}{\sqrt{2}}
    \begin{bmatrix}
        x_{11} + x_{(n+1)1} & x_{12}+x_{(n+1)2}  & \cdots & x_{1(m)} + x_{(n+1)(m)} \\ 
        x_{21} + x_{(n+2)1} & x_{22}+x_{(n+2)2}  & \cdots & x_{2(m)} + x_{(n+2)(m)} \\
        \vdots & \vdots  & \cdots & \vdots \\ 
        x_{(n)1} + x_{(2n)1}  & x_{(n)2} + x_{(2n)2} & \cdots & x_{(n)m} + x_{(2n)m}  \\
    \end{bmatrix}\\
    Y^T &=& \frac{1}{\sqrt{2}}
    \begin{bmatrix}
        x_{11} + x_{(n+1)1} & x_{21}+x_{(n+2)1}  & \cdots & x_{(n)1} + x_{(2n)1} \\ 
        x_{12} + x_{(n+1)2} & x_{22}+x_{(n+2)2}  & \cdots & x_{(n)2} + x_{(2n)2} \\
        \vdots & \vdots  & \cdots & \vdots \\ 
        x_{1(m)} + x_{(n+1)(m)}  & x_{2(m)} + x_{(n+2)(m)} & \cdots & x_{(n)m} + x_{(2n)m}  \\
    \end{bmatrix}
  \end{eqnarray}
\end{widetext}

\begin{widetext}
  \begin{eqnarray}
    &&YY^T \in R^{n\times n} \\
    &&=\frac{1}{2}
    \resizebox{0.9\hsize}{!}
    {$\begin{bmatrix}
        M_{11} + M_{1(n+1)}+ M_{(n+1)1}+M_{(n+1)(n+1)} & M_{12} + M_{1(n+2)} + M_{(n+1)2}+ M_{(n+1)(n+2)} & \cdots & M_{1(n)} + M_{1(2n)} + M_{(n+1)(n)}+ M_{(n+1)(2n)}\\
        M_{21} + M_{2(n+1)} + M_{(n+2)1}+ M_{(n+2)(n+1)}  & M_{22} + M_{2(n+2)} + M_{(n+2)2}+ M_{(n+2)(n+2)} & \cdots & M_{2(n)} + M_{2(2n)} + M_{(n+2)(n)}+ M_{(n+2)(2n)} \\
        \vdots \\
        M_{(n)1} + M_{(n)(n+1)} + M_{(2n)1}+ M_{(2n)(n+1)} & M_{(n)2} + M_{(n)(n+2)} + M_{(2n)2}+ M_{(2n)(2+n)} & \cdots & M_{(n)(n)} + M_{(n)(2n)} + M_{(2n)(n)}+ M_{(2n)(2n)}
    \end{bmatrix}$}
    \nonumber \\
    &&=\frac{1}{2}
    \resizebox{0.9\hsize}{!}
    {$\begin{bmatrix}
        2 + 2M_{1(n+1)} & M_{12} + M_{1(n+2)} + M_{(n+1)2}+ M_{(n+1)(n+2)} & \cdots & M_{1(n)} + M_{1(2n)} + M_{(n+1)(n)}+ M_{(n+1)(2n)}\\
        M_{21} + M_{2(n+1)} + M_{(n+2)1}+ M_{(n+2)(n+1)}  & 2 + 2M_{2(n+2)} & \cdots & M_{2(n)} + M_{2(2n)} + M_{(n+2)(n)}+ M_{(n+2)(2n)} \\
        \vdots \\
        M_{(n)1} + M_{(n)(n+1)} + M_{(2n)1}+ M_{(2n)(n+1)} & M_{(n)2} + M_{(n)(n+2)} + M_{(2n)2}+ M_{(2n)(2+n)} & \cdots & 2 + 2M_{(n)(2n)}
    \end{bmatrix}$}
    \nonumber \\
    && = I+\frac{1}{2}\left(A+B\right)  
    \nonumber 
  \end{eqnarray}
\end{widetext}

\begin{widetext}
  \begin{eqnarray}
    && A = 
    \resizebox{0.9\hsize}{!}
    {$\begin{bmatrix}
        M_{1(n+1)} & M_{12} + M_{(n+1)2} & \cdots & M_{1(n)} + M_{(n+1)(n)}\\
        M_{21} + M_{(n+2)1} & M_{2(n+2)} & \cdots & M_{2(n)} + M_{(n+2)(n)} \\
        \vdots \\
        M_{(n)1} + M_{(2n)1} & M_{(n)2} + M_{(2n)2} & \cdots & M_{(n)(2n)}
    \end{bmatrix}
    B = 
    \begin{bmatrix}
        M_{1(n+1)} & M_{1(n+2)} + M_{(n+1)(n+2)} & \cdots & M_{1(2n)}+ M_{(n+1)(2n)}\\
        M_{2(n+1)} + M_{(n+2)(n+1)}  & M_{2(n+2)} & \cdots & M_{2(2n)} + M_{(n+2)(2n)} \\
        \vdots \\
        M_{(n)(n+1)} + M_{(2n)(n+1)} & M_{(n)(n+2)} + M_{(2n)(2+n)} & \cdots & M_{(n)(2n)}
    \end{bmatrix}$}
    \nonumber 
  \end{eqnarray}
\end{widetext}

\begin{widetext}
  \begin{eqnarray}
    && A^T = 
    \resizebox{0.9\hsize}{!}
    {$\begin{bmatrix}
        M_{1(n+1)} & M_{21} + M_{(n+2)1} & \cdots & M_{(n)1} + M_{(2n)1}\\
        M_{12} + M_{(n+1)2} & M_{2(n+2)} & \cdots & M_{(n)2} + M_{(2n)2} \\
        \vdots \\
        M_{1(n)} + M_{(n+1)(n)} & M_{2(n)} + M_{(n+2)(n)} & \cdots & M_{(n)(2n)}
    \end{bmatrix}
    B^T = 
    \begin{bmatrix}
        M_{1(n+1)} & M_{2(n+1)} + M_{(n+2)(n+1)} & \cdots & M_{(n)(n+1)} + M_{(2n)(n+1)}\\
        M_{1(n+2)} + M_{(n+1)(n+2)}  & M_{2(n+2)} & \cdots &  M_{(n)(n+2)} + M_{(2n)(2+n)} \\
        \vdots \\
        M_{1(2n)}+ M_{(n+1)(2n)} &M_{2(2n)} + M_{(n+2)(2n)} & \cdots & M_{(n)(2n)}
    \end{bmatrix}$}
    \nonumber \\
  \end{eqnarray}
\end{widetext}

\begin{widetext}
  \begin{eqnarray}
    && YX^T \nonumber \\
    && = \frac{1}{\sqrt{2}}
    \begin{bmatrix}
        M_{11} + M_{(n+1)1} & M_{12}+M_{(n+1)2} & \cdots & M_{1(2n)} + M_{(n+1)(2n)} \nonumber \\
        M_{21} + M_{(n+2)1} & M_{22}+M_{(n+2)2} & \cdots & M_{2(2n)} + M_{(n+2)(2n)} \nonumber \\
        \vdots & \vdots & \cdots & \vdots \nonumber \\
        M_{(n)1} + M_{(2n)1} & M_{n2}+M_{(2n)2} & \cdots & M_{n(2n)} + M_{(2n)(2n)} \nonumber
    \end{bmatrix} \nonumber \\
     &&= \frac{1}{\sqrt{2}}
     \begin{bmatrix}
        1 + M_{1(n+1)} & M_{12}+M_{(n+1)2} & \cdots & M_{1(2n)} + M_{(n+1)(2n)} \nonumber \\
        M_{21} + M_{(n+2)1} & 1+M_{2(n+2)} & \cdots & M_{2(2n)} + M_{(n+2)(2n)} \nonumber \\
        \vdots & \vdots & \cdots & \vdots \nonumber \\
        M_{(n)1} + M_{(2n)1} & M_{n2}+M_{(2n)2} & \cdots & 1+M_{n(2n)} \nonumber
    \end{bmatrix}
    = \frac{1}{\sqrt{2}}\left[I + A \vert I+B \right] \nonumber \\
    &&XY^T = \frac{1}{\sqrt{2}}
    \begin{bmatrix}
        1 + M_{1(n+1)} & M_{21}+M_{(n+2)1} & \cdots & M_{(n)1} + M_{(2n)1} \nonumber \\
        M_{12} + M_{(n+1)2} & 1+M_{2(n+2)} & \cdots & M_{n2} + M_{(2n)2} \nonumber \\
        \vdots & \vdots & \cdots & \vdots \nonumber \\
        M_{1(2n)} + M_{(n+1)(2n)} & M_{2(2n)}+M_{(n+2)(2n)2} & \cdots & 1+M_{n(2n)} \nonumber
    \end{bmatrix}
    = \frac{1}{\sqrt{2}}\left[\frac{I+A^T}{I+B^T} \right] \nonumber 
  \end{eqnarray}
\end{widetext}

The projection error is,
\begin{eqnarray} \label{projection_error}
    \mathcal{F} &=& XX^T - XY^T(YY^T)^{-1} Y Y^T (YY^T)^{-1}Y X^T \nonumber \\ 
    &=& XX^T - XY^T (YY^T)^{-1}YX^T.
\end{eqnarray}

If the off-diagonal terms are small, we can expand the inverse using Neuman series~\cite{petersen2008matrix}. 
\begin{eqnarray}
    \mathcal{E} &=& Tr\left[XX^T - XY^T(YY^T)^{-1}Y X^T\right] \nonumber \\ 
    &=& 2n - Tr\left[XY^T (YY^T)^{-1}YX^T\right]
\end{eqnarray}

To the first order,  
\begin{widetext}
  \begin{eqnarray}
  XY^T(YY^T)^{-1}YX^T \simeq 
    \begin{bmatrix}
        2 +2 M_{1(n+1)} & 0 & \cdots & 0\\
        0 & 2 + 2M_{2(n+2)} & \cdots & 0 \\
        \vdots \\
        0& 0 & \cdots & 2  + M_{(n)(2n)}
    \end{bmatrix}
  \end{eqnarray}
\end{widetext}
and the error is $\Delta\mathcal{E}\simeq -2\sum_{i} M_{(i)(n+i)}$. 
This result teaches us that the projection error is minimized by maximizing the pair correlations to the first-order correction.

\underline{Second order in consideration}
The first order results do not involve any correlation between the coarse-grained pair and the rest, which is captured in the second order terms.

The inverse term is expanded as,
\begin{eqnarray}
    \left(YY^T\right)^{-1} &\simeq&  I-\frac{1}{2}\left(A+B\right) + \frac{1}{4}\left(A+B\right)^2 + \mathcal{O}\left(M^3\right) \nonumber\\
    YX^T &=& I + A + B
\end{eqnarray}
Then, 
\begin{widetext}
  \begin{eqnarray}
  XY^T\left(YY^T\right)^{-1}YX^T =
  \resizebox{0.75\hsize}{!}
    {$
    \begin{bmatrix}
        I+\frac{1}{2}\left(A+B\right)-\frac{1}{4}\left(A^2+B^2+AB+BA\right) & \cdots \\ \cdots & I+\frac{1}{2}\left(A+B\right)-\frac{1}{4}\left(A^2+B^2+AB+BA\right) 
    \end{bmatrix}$} \nonumber
  \end{eqnarray}
\end{widetext}

\begin{widetext}
  \begin{eqnarray}
    \Delta\mathcal{E}\simeq \sum_{i} -2M_{(i)(N+i)}+2\sum_{i=1}^{N}\sum_{j=1,j\ne i}^{N}\left(M_{ij}M_{i(N+j)}+M_{(N+i)j}M_{(N+i)(N+j)} \right)
  \end{eqnarray}
\end{widetext}

So, the error is minimized by minimizing the correlation between the blocks of coarse-graining matrix. 
This can be regarded an information loss of data structure during coarse-graining. 
We reconsider the coarse-graining process in this aspect.

\subsection{Computer implementation: Stepwise Coarse-Graining}

We repeat the projection error calculation during the coarse-graining process in a stepwise manner. Starting with correlation matrix $M$,
we apply the coarse-graining process as the updates of the row and column of coarse-graining data. 
For instance, we coarse-grain data $1$ and $2$, then, the correlation matrix will be updated a following matrices.
\begin{widetext}
  \begin{eqnarray}
    \mathbf{u_1}=-
    \begin{bmatrix}
        1 & 0 \\ 0 & M_{21} \\ 0 & M_{31} \\
        \vdots & \vdots \\ 0 & M_{N1}
    \end{bmatrix}, 
    \mathbf{v_1}=
    \begin{bmatrix}
        0 & 1 \\ M_{21} & 0 \\ M_{31} & 0 \\
        \vdots & \vdots \\ M_{N1} & 0
    \end{bmatrix},
        \mathbf{u_2}=
    \begin{bmatrix}
        0 & 0 \\ 1 & 0 \\ 0 & M_{31} \\
        \vdots & \vdots \\ 0 & M_{N1}
    \end{bmatrix}, 
    \mathbf{v_2}=
    \begin{bmatrix}
        0 & 0 \\ 0 & 1 \\ M_{31} & 0 \\
        \vdots & \vdots \\ M_{N1} & 0
    \end{bmatrix}
  \end{eqnarray}
\end{widetext}
The stepwise coarse-graining is represented as $M' = M + \mathbf{u_1}\mathbf{v_1}^T + \mathbf{u_2}\mathbf{v_2}^T$. 
The projection error involves the inverse of the coarse-grained matrix $M'$, which can be calculated using the Woodbury identity~\cite{petersen2008matrix}.

In the limit that $M_{ij}^{-1} \gg \frac{M_{i1}^{-1}M_{1j}^{-1}}{M_{11}^{-1}}$,
after some algebra, the error is expressed as \begin{eqnarray}
    M_1RM_1^T \propto \frac{1+2M_{12}}{1+2\sum_{k=3}^{N}M_{1k}M_{k2}^{-1}}
\end{eqnarray}
If we further assume that $M_{k2}^{-1}\simeq -M_{k2}$, 
\begin{eqnarray}
    M_1RM_1^T \propto \frac{1+2M_{12}}{1-2\sum_{k=3}^{N}M_{1k}M_{k2}}.
\end{eqnarray}
Thus, the optimal pair should maximize their correlation as well as the correlations between pair and the rest.

Here, we describe our computer implementation of the stepwise coarse-graining procedure, which we have linked to in the introduction (todo).  Our code makes use of the the linselect class, GenSelect, which is a flexible variant of the stepwise linear selection algorithm.  Primarily this is useful here because it contains all the code we need to carry out Woodbury updates \cite{petersen2008matrix} to the inverse correlation matrix, etc., of the set of features we will retain in our coarse-grained model.

The idea behind the linear stepwise algorithm is as follows:  We start in a given stage with a set of $2n$ base features, and we work towards grouping these into $n$ pairs, each an average of two of the original features.  The goal is again to find a set of $n$ pairs that together allow us to accurately predict the behavior of all the original $2n$ features via linear projection.  As noted above, the full set of possible pairings is very large, preventing an exhaustive search for the best possible set of $n$ pairs.  However, we can use a step-wise procedure to find a relatively good set of pairings:   We first identify which single feature between $2$ and $2n$ should be paired with feature $1$ to obtain a single-feature model that minimizes projection error (\ref{projection_error}) summed over the original $2n$ features.  The pair-averaged feature chosen is added to our ``coarse-grained model" feature set, $s$. We then proceed iteratively from there:  If feature $2$ was not paired with feature $1$ in the first step, we ask which as-yet-unpaired feature it should be paired with, such that when the pair is added to $s$, the net projection error on the original features is reduced as much as possible, etc.

The key measure that we need to consider to decide which feature to pair with a given starting feature is the decrease in projection error of the original features when they are mapped to the span of the features in $s$.  We want to choose the pair that maximizes this decrease.  This can be read out from (18) of \cite{landy2017stepwise}:  When feature $j$ is moved to $s$, the projection error of feature $i$ moves to
\begin{eqnarray}\label{component_projection_error}
    \mathcal{F}_i \leftarrow \mathcal{F}_i - \frac{\vert M_{ij} - U_{ij} \vert^2}{1 - U_{jj}}.
\end{eqnarray}
Here, $M$ is again the correlation matrix and $U$ is a second accounting matrix that is tracked throughout the linear selection process.  In our application, we sum over $i$ in this line over all the original $2n$ features to get the projection error improvement of the original features, and $j$ will be a candidate pair feature.  Conveniently, the $M$ and $U$ values for a pair feature can be easily obtained from those of the original features:  E.g., if we let $1$ and $2$ be two of the original features that we're considering pairing and adding to $s$, we have
\begin{equation}
    M_{i(12)} = \frac{1}{\sqrt{2(1 + M_{12}})} \left (M_{i1} + M_{i2} \right),
\end{equation}
where we use $(12)$ as the index for the pair.  This equation allows us to read out the correlation matrix values for any pair from those for the original features.  A similar identify holds for the  values of $U$ along pairs not yet added to $s$.  With these two results, we can easily evaluate (\ref{component_projection_error}) to identify the benefit of adding a particular pair.  In a given step, then, we simply identify the potential gain of each potential pair for a given feature, find the best option, and then add it to $s$ using linselect.

The computational complexity of this process is as follows:  (1) Construction of the base correlation matrix requires $O(n^2 m)$ computations, where $m$ is the number of data samples we have.  We require $m \geq 2n$ to obtain a fully linearly independent set of base features, so this  is at least $O(n^3)$.  (2) To evaluate the projection error improvement for a given pair (\ref{component_projection_error}), we need to construct the $M$ and $U$ values for that pair along both the base features as well as any pair that has already been added to $s$.  This requires $O(n)$ computations. However, there are $O(n)$ possible pairs to consider in a given step, so in total it takes $O(n^2)$ computations to identify the scores for each possible pair in a given step.  Selecting the best among these takes only $O(n)$ computations.  (3)  Adding the best possible pair to $s$ requires some calculations to be carried out within linselect.  This takes $O(n^2)$ computations as well \cite{landy2017stepwise}.  Points (2) and (3) together imply that each step requires $O(n^2)$ effort to carry out.  There are $n$ steps total, so the coarse-graining requires $O(n^3)$ computations total, the same scaling as is required to construct the original correlation matrix.

\section{Extension to the Non-linear analysis}
In the previous section, we learned that the direct measurement of the change in mutual information during the coarse-graining step is ill-defined. Instead, we can use a greedy optimization approach to minimize the mutual information change during the stepwise coarse-graining process. This approach extends the linear version of the stepwise coarse-graining process to a non-linear one. Although this optimization may not be global, it can still provide a meaningful approximation. 

In stepwise coarse-graining, we coarse-grain two input data minimizing the information loss. If we label $X$ to represent the two input data, and the remaining input data are labeled $X^c$. The coarse-grained data is labeled $Y$. Since the correlation between two Gaussian-like datasets corresponds to the mutual information of general distribution, it is rational to use mutual information instead of correlation for non-linear analysis.

Thus, the condition that maximizes the correlation of the selected pairs $X$ corresponds to the maximal compression of the data $X$ to $Y$, which minimizes their mutual information. (The whole single data is compressed!) On the other hand, the condition that maximizes the correlation between the selected pair and the remaining data corresponds to maximizing the mutual information between $X^c$ and $Y$.

Our goal is to compress the data as much as possible while preserving the correlated structure of the coarse-grained data. 
It will extract mesoscopic collective properties discarding irrelevant information, which is consistent with the information bottleneck idea~\cite{tishby2000information,tishby2015deep}. 

\begin{eqnarray}
    \min_{Pairs\left[x_1, x_2, \cdots, x_{2N}\right]} I(X;Y) - \beta I(X^c;Y).
\end{eqnarray}
where $Pairs\left[x_1, x_2, \cdots, x_{2N}\right]$ is the possible pair combinations, such as $\left\{\left(x_1,x_{N+1}\right), \left(x_2,x_{N+2}\right),\cdots,\left(x_N,x_{2N}\right)\right\}$.
$I(X;T)$ corresponds to the information loss of grouping two nodes into one. Note that the delta function does not appear during the stepwise process.
$I(T;Y)$ measures the information decreasing during transmission. 

\section{Results}

\subsection{Mean Variance function}
Meshulam and Bialek showed that the mean-variance of the data follows the power law~\cite{meshulam2017collective,meshulam2019coarse}
\begin{eqnarray}
    M_2\left(k\right)\equiv \frac{1}{N_k}\sum_{i=1}^{N_k}\left[\left\langle \left(X_i^{k}\right)^2\right\rangle - \left\langle X_i^{k}\right\rangle^2\right].
\end{eqnarray}
If data are from Gaussian, the mean-variance is proportional to $k$; hence the exponent in the power law would be $1$. On the other hand, if they are all correlated, the exponent will be $2$. 
Any finite correlation will give intermediate values for the exponent. 

\subsection{Activity function}
The probability of the coarse-grained data is expressed as,
\begin{eqnarray}
    P_K\left(X^{(K)}\right)=P_{0}\left(K\right)\delta\left(X^{(K)}\right) + \left[1-P_0\left(K\right) \right] Q_K\left(X^{(K)}\right)
\end{eqnarray}
where $K = 2^{k-1}$. $P_{0}$ is a probability that all $K$ original data in the coarse-grained cluster of size $K$ are zero. 
Note that random data $P_{0}$ satisfies the recurrence relation $P_{0}\left(K\right) = P_{0}\left(K-1\right)\times P_{0}\left(K-1\right)$ which yields  $P_{0}\left(K\right) = \exp\left(-a K\right)$.
In the presence of the collective behavior, the decay is slower than the random process suggesting 
$P_{0}\left(K\right) = \exp\left(-a K^\beta \right)$ with $\beta <1$.

\subsection{Rank and MSE}
\begin{figure}[h!]
\centering
\includegraphics[width=8cm]{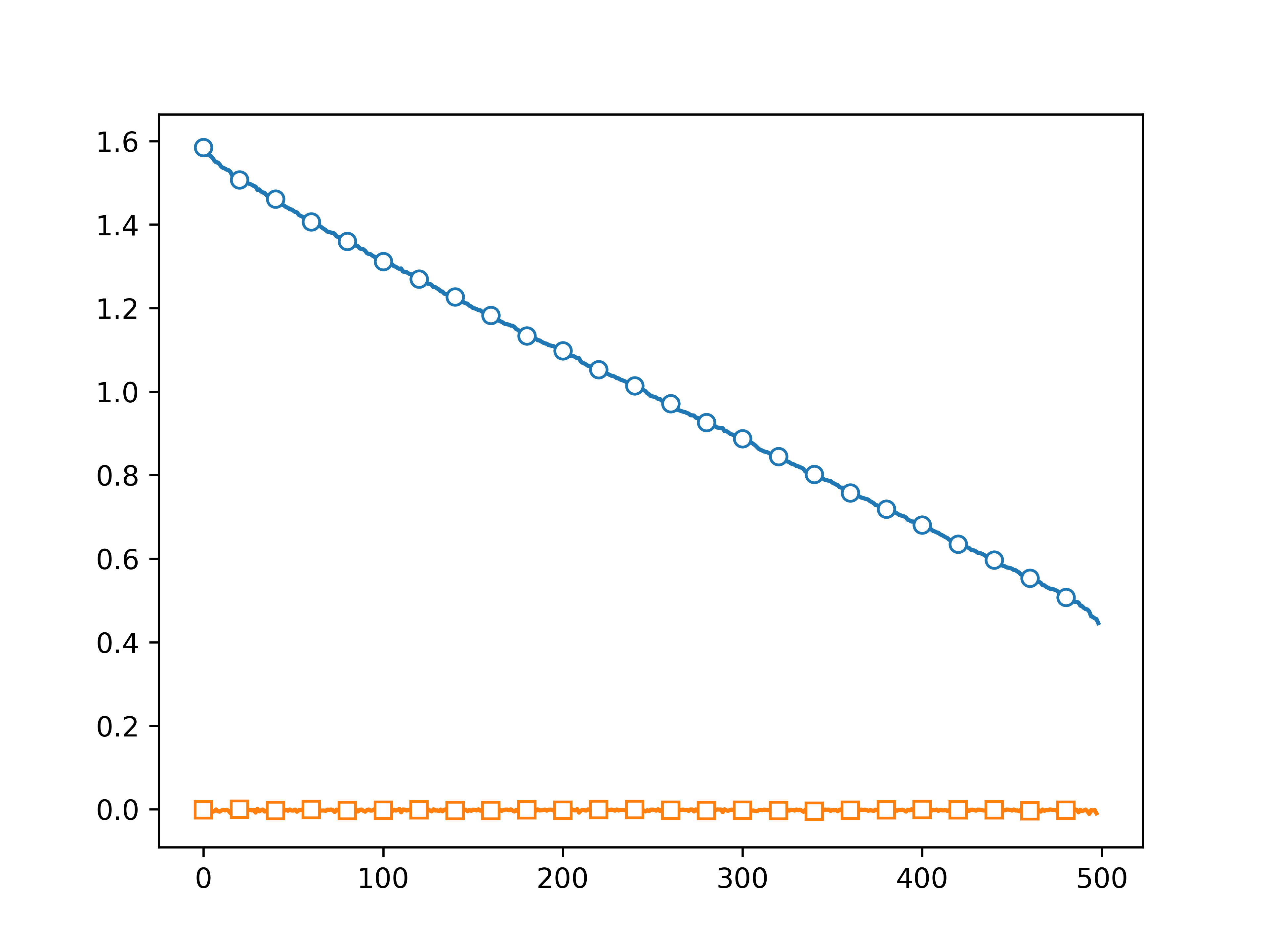}
\caption{Random Matrix: The COD of the random matrix decreases linearly to rank. Hence, the increment is almost constant.}
\label{fig:RM}
\end{figure}

In this subsection, we will discuss how the coefficient of determination distributes to the rank of PCA or the linselect. It is suggested that the eigenvalues follow the power law to the rank for PCA. Here we will show that similar relation holds for the linselect process. 

Feature selection process yields mean squared error (MSE), 
\begin{eqnarray}
    \mathcal{F}_i &=& XX^T - X\left\{ P^{(i)}+P^{(i)T}\right\}X^T - XP^{(i)}P^{(i)T}X^T \nonumber \\
    &=& XX^T - X\tilde{X}^T\left(\tilde{X}\tilde{X}^T\right)^{-1} \tilde{X}X^T = \sum_{j} C(i,j).
\end{eqnarray}
$P^{(i)}$ is the projection error of $i$ row removed. 

In case of random distribution, the pair correlation might be considered as uniform. After $i^{th}$ deletion step of linselect, the density of the data decreases by $\frac{N-i}{N}$. 
\begin{eqnarray}
    \int_{0}^{r_N} dr r^{d-1} \rho_i  C\left(i,j\right)(r) &=& \frac{N-i}{V}\int_{0}^{r_N} dr r^{d-1} C\left(i,j\right)(r) \nonumber \\
    &\propto& \left(N-i\right).
\end{eqnarray}
Figure~\ref{fig:RM} shows that the increment of the coefficient of determination in the linselect is almost constant, which is consistent as we expected.

In the vicinity of critical point,  $C\left(i,j\right)(r)\propto \frac{1}{r^{d-2+\eta}}$. At $N-i$, the average distance is rescaled by $x = r\xi$ where $\xi = \left(\frac{N}{N-i}\right)^{\frac{1}{d}}$. 
\begin{eqnarray}
    \int_{0}^{r_N} dx x^{d-1} C\left(i,j\right)(x) &\propto& \int_{0}^{r_N} dx x^{d-1} {x^{-d+2-\eta}} \nonumber \\
    &=& \int_{0}^{r_N} dr r^{1-\eta} \xi^{2-\eta} \nonumber \\
    &\propto& \left( N-i\right)^{\frac{\eta-2}{d}} C_0
\end{eqnarray}
where $C_0$ is the largest COD. 
Thus, in the vicinity of the critical point, the cod follows a power-law decay. 

By judging if the coefficient of determination obeys the random distribution, we may identify the collective behaviors. 

\section{Application}
\begin{figure*}
    \centering
    \begin{subfigure}[t]{0.3\textwidth}
        \includegraphics[width=\textwidth]{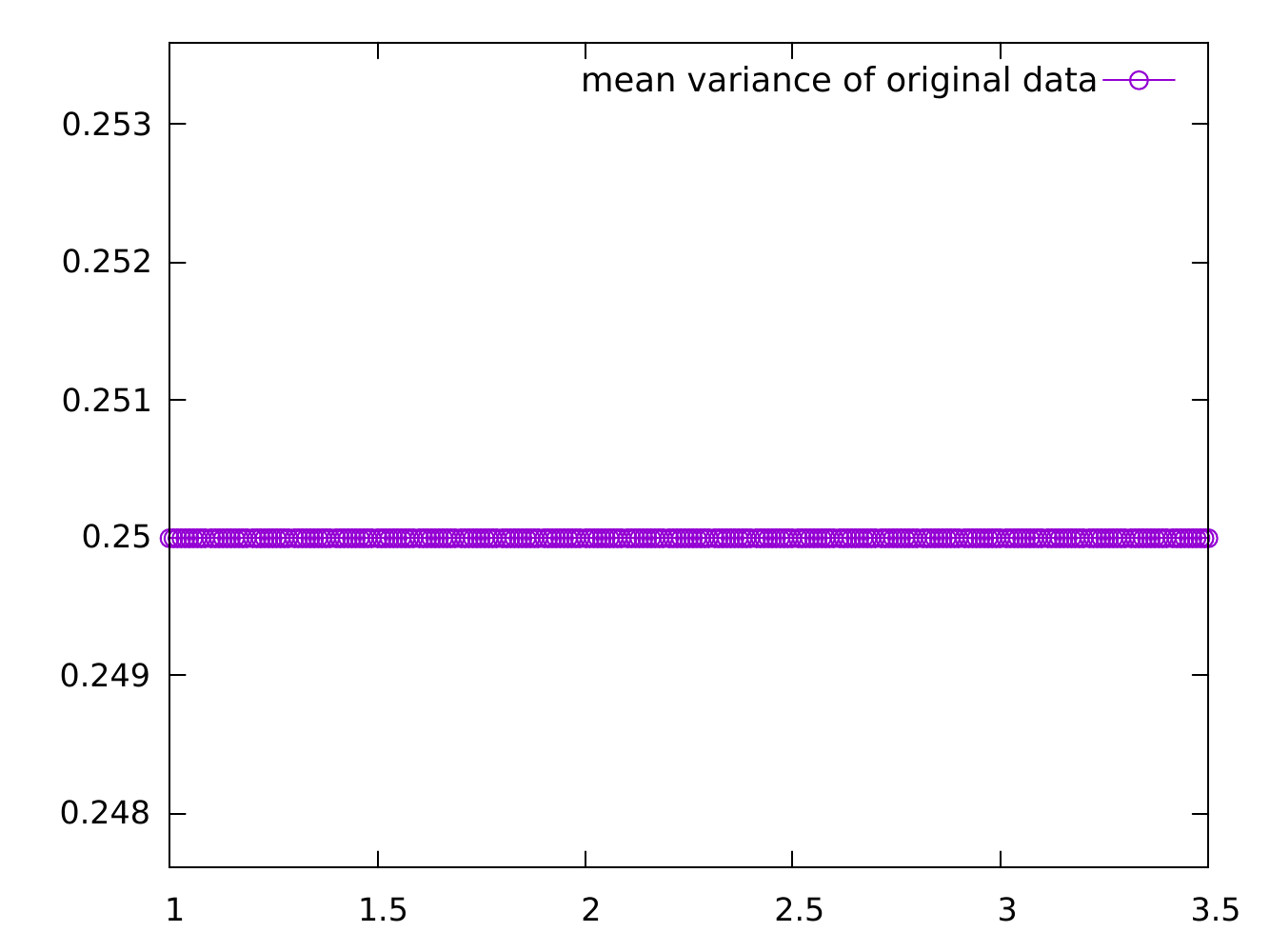}
        \caption{Mean-variance of original data}
        \label{fig:Ising-a}
    \end{subfigure}
    \begin{subfigure}[t]{0.3\textwidth}
        \includegraphics[width=\textwidth]{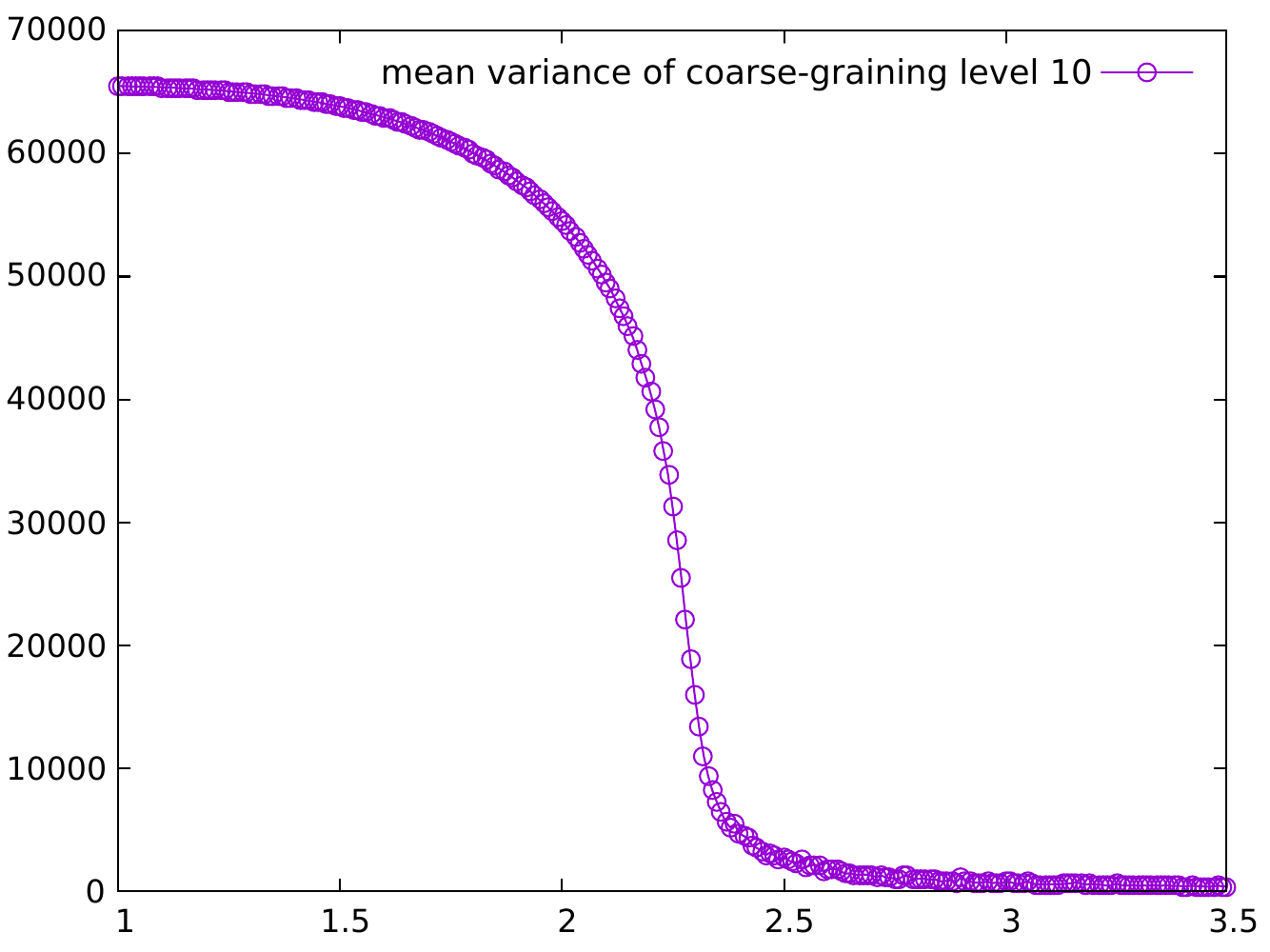}
        \caption{Mean-variance at coarse-graining level 10}
        \label{fig:Ising-b}
    \end{subfigure}
    \begin{subfigure}[t]{0.3\textwidth}
        \includegraphics[width=\textwidth]{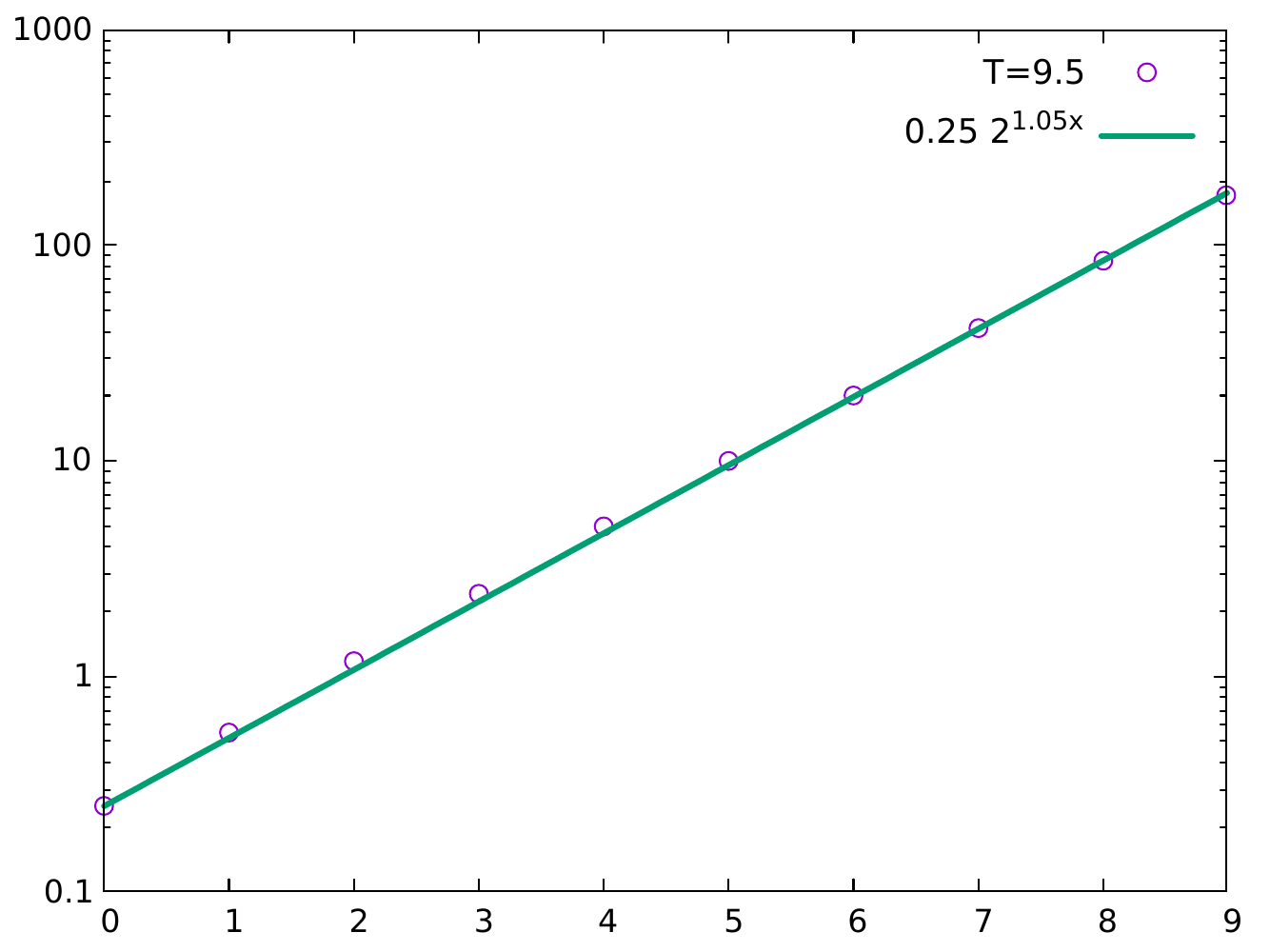}
        \caption{Mean-variance vs coarse-graining level at $T=9.5$}
        \label{fig:Ising-c}
    \end{subfigure}
    \begin{subfigure}[t]{0.3\textwidth}
        \includegraphics[width=\textwidth]{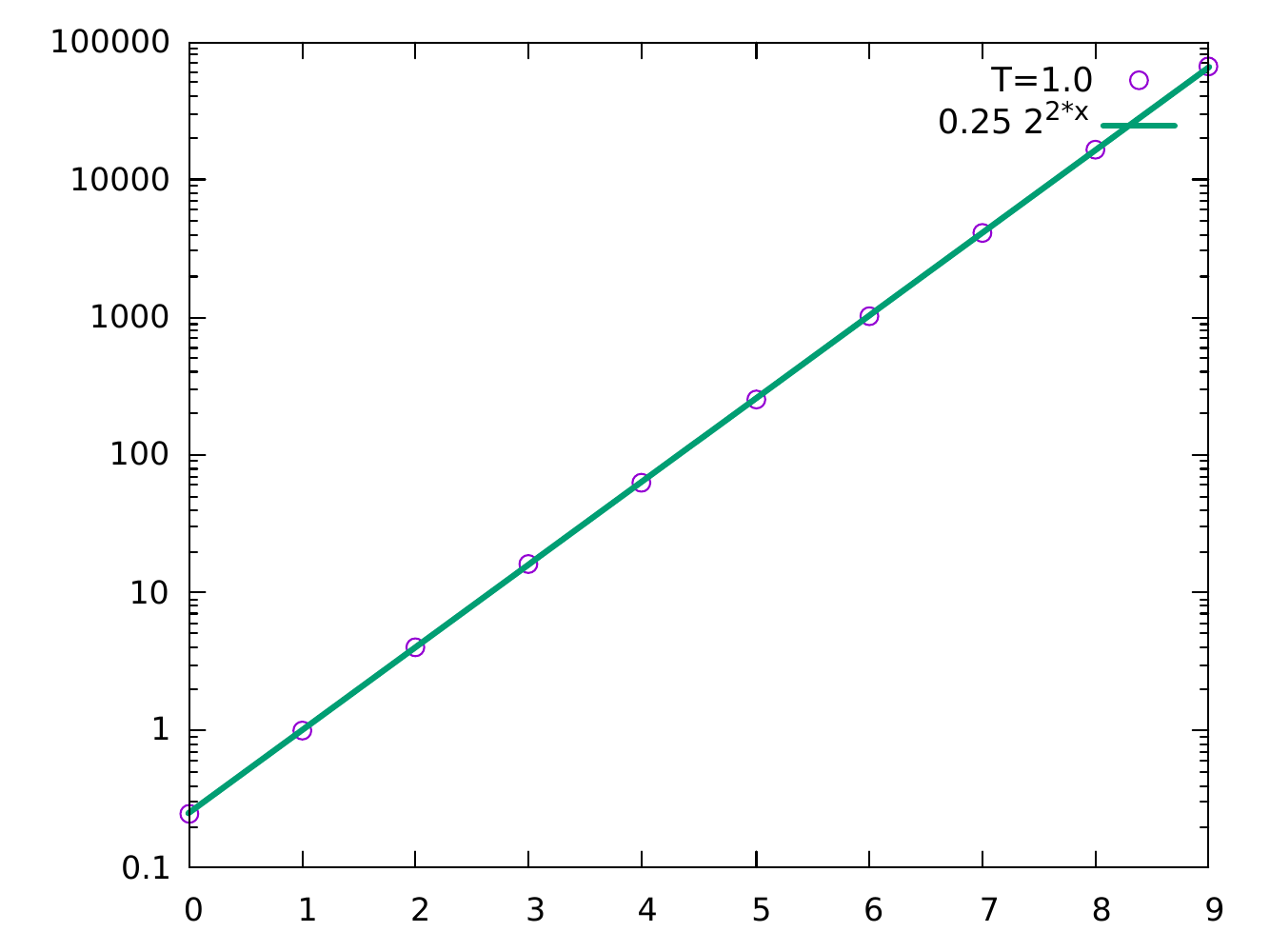}
        \caption{Mean-variance vs coarse-graining level at $T=1.0$}
        \label{fig:Ising-d}
    \end{subfigure}
    \begin{subfigure}[t]{0.3\textwidth}
        \includegraphics[width=\textwidth]{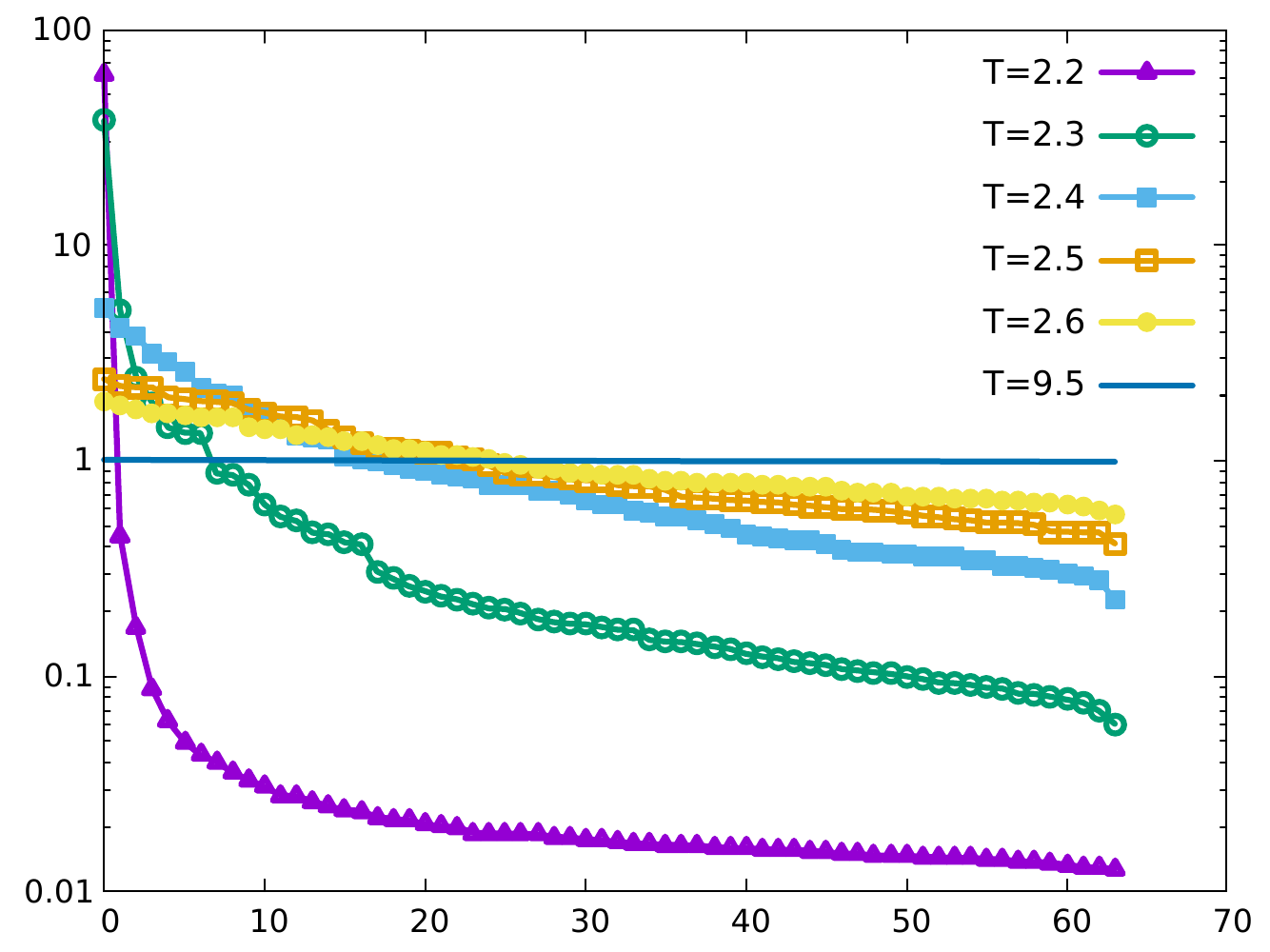}
        \caption{COD with respect to rank at different temperatures}
        \label{fig:Ising-e}
    \end{subfigure}
    \begin{subfigure}[t]{0.3\textwidth}
        \includegraphics[width=\textwidth]{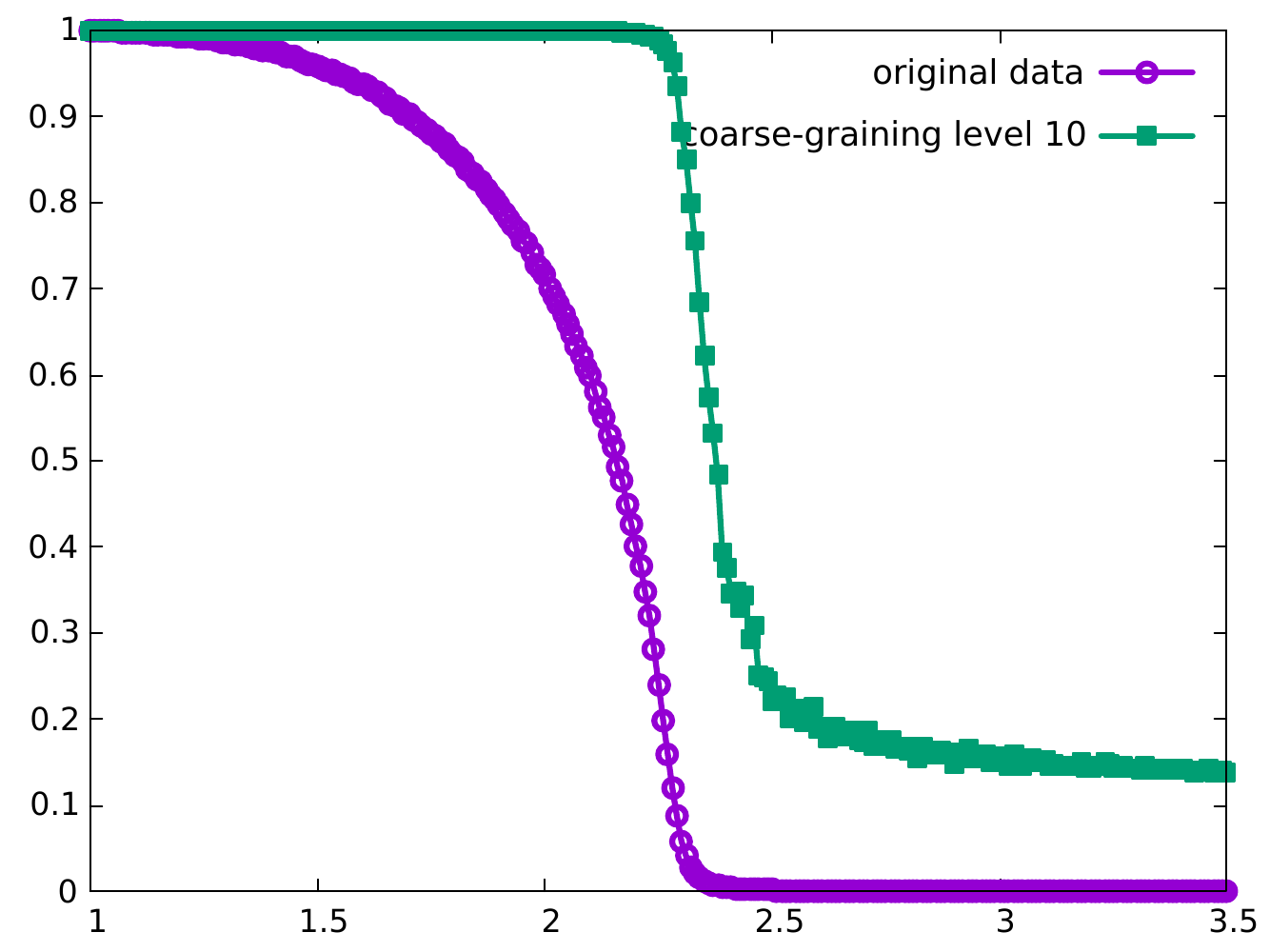}
        \caption{Normalized COD at different coarse-graining levels}
        \label{fig:Ising-f}
    \end{subfigure}
    \caption{Our model is tested to Ising model.}
\end{figure*}

In Figure~\ref{fig:RM}, the random matrix is tested. We generate a matrix $X$ with randomly sampled elements from the normal distribution function of mean $0$ and variance $1$.
The MSE linearly decreases as the rank increases, and the increment is nearly uniform, consistent with our previous section's claim. 
The trend becomes more apparent and conspicuous as the coarse-graining step is repeated. 
Later we will see that some nonlinear COD distribution converges to a uniform increment after several coarse-grained steps implying that the correlated features disappear over finite coarse-grained size. Thus, this can serve as a reference value for correlated data. 

Next, we apply our approach to the 2D Ising model with a lattice size of $64\times 64$. 
It is better for visualizing the mean-variance setting the rescaling constant as $1$. 
Figures~\ref{fig:Ising-a} and \ref{fig:Ising-b} show the mean-variance of the original data and the coarse-grained data at level $10$. 
The original data's mean-variance data is $0.25$ because the probabilities for up and down spin are the same. 
Interestingly, the transition is revealed after coarse-graining steps (\ref{fig:Ising-b}). 
The mean-variance grows according to the power-law as coarse-graining steps increase. The exponent of the power-law is almost $1$ at high temperatures. For example, it is $1.05$ at $T=9.5$~\ref{fig:Ising-c}. At high temperatures, the states are random, and the data distribution follows Gaussian. On the other hand, the exponent jumps to values close to $2$ below the critical point. At low temperatures, the correlation of Ising spins extends to infinity, which yields the exponent $2$~\ref{fig:Ising-d}, consistent with our claim.

Figure~\ref{fig:Ising-e} shows the COD of linselect. At high temperatures $T=9.5$, the COD is uniform over the rank, just like random data. Below the critical point, low rank COD dominates, and COD sharply drops. At intermediate states, low ranks are non-linear while high ranks are linear. The trend changes substantially between $T=2.3$ and $T=2.4$.
The non-linear behavior in low ranks dims as coarse-graining steps are repeated if the temperature is above the critical point. On the other hand, it is still persistent if the temperature is below the critical point. It provides an estimate of an average correlation size at transient temperatures. 

The coarse-graining steps accentuate this tendency. Figure~\ref{fig:Ising-f} shows that the sum of the largest normalized COD changes sharply at the critical point. 
The largest normalized COD of the 10th coarse-grained data shows a much narrower transition than that of the original data. The intermediate states in the original data move to one of the states as coarse-graining step repeats, similar to a conventional renormalization group process.
It justifies the usefulness of the RG-like approach to data analysis.

We apply our method to the analysis of the glass simulation. The position of the glass particle changes during the simulation, so the conventional RG methods applied to a lattice system are not valid. 
To apply our approach, we need time-series data. Their correlation will measure the proximity of the data. For this purpose, we borrow the measure for the particle exchange event, $p_{hop}$, which encodes the kinetic properties of the glass particles. 
The formal definition is as follows:
\begin{eqnarray}
    p_{hop} \equiv \sqrt{\left\langle \left(r_i - \langle r_i \rangle_B\right)^2 \right\rangle_A \left\langle \left(r_i - \langle r_i \rangle_A\right)^2 \right\rangle_B}
\end{eqnarray}
where $A=\left[t-t_R/2,t\right],B=\left[t,t+t_R/2\right]$. We choose $t_R$ as $10 \tau$.
When a glass particle is in a local cage surrounded by its neighbors, the value of $P_{hop}$ remains small. However, it grows while it exchanges its position with one of its neighbors, overcoming the kinetic energy barrier. If there is a chain of exchange events, perhaps triggered by one large exchange event inducing subsequent local responses of neighboring particles~\cite{lee2020local}, $P_{hop}$ is locally correlated. If a finite range correlation exists, the correlation can propagate. This measure can be used to identify localized population dynamics in crowded systems. 

\begin{figure*}
    \centering
    \begin{subfigure}[b]{0.3\textwidth}
        \includegraphics[width=\textwidth]{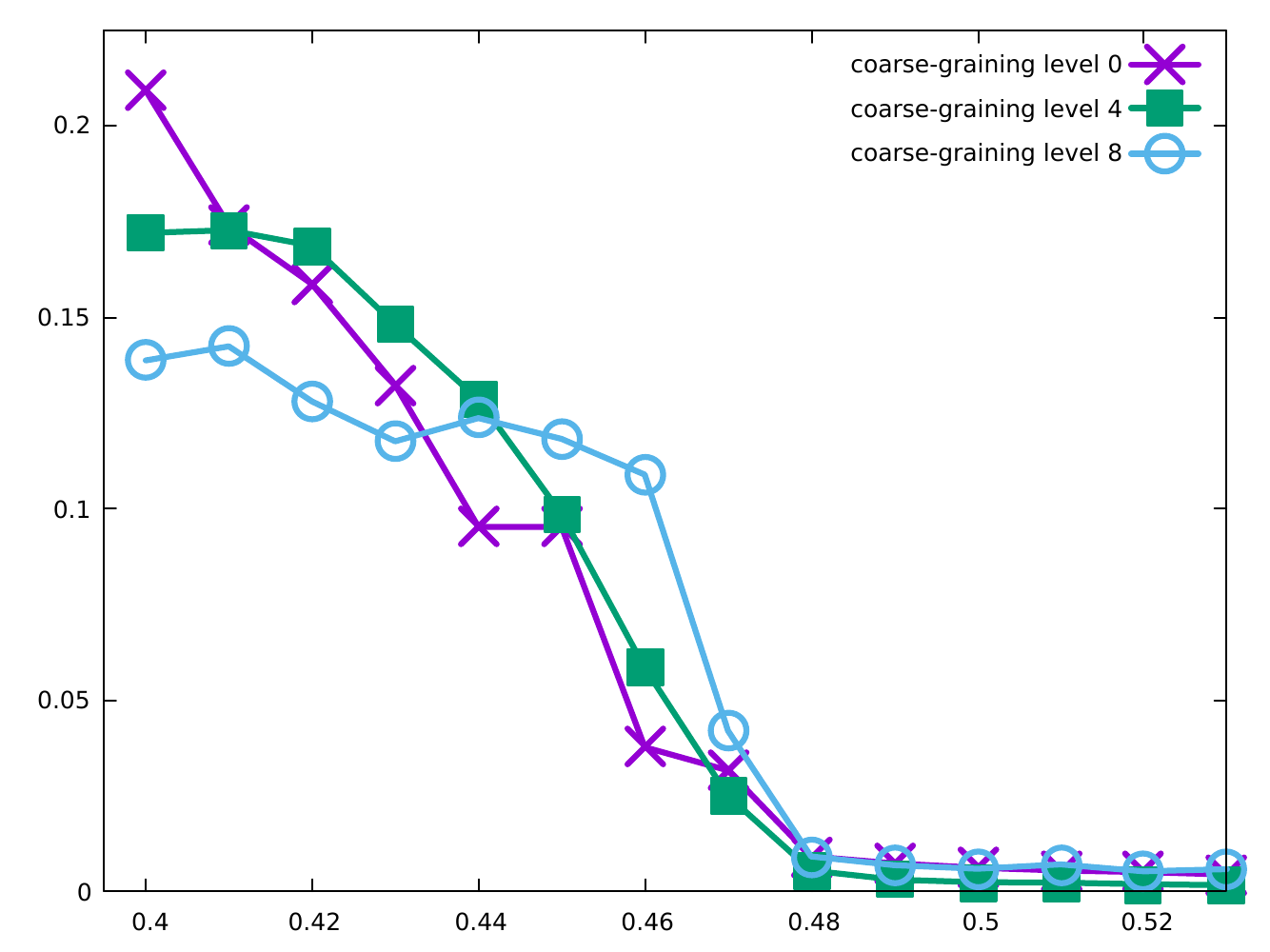}
        \caption{transient behavior of COD in Ionic liquid}
        \label{fig:IL-a}
    \end{subfigure}
    \begin{subfigure}[b]{0.3\textwidth}
        \includegraphics[width=\textwidth]{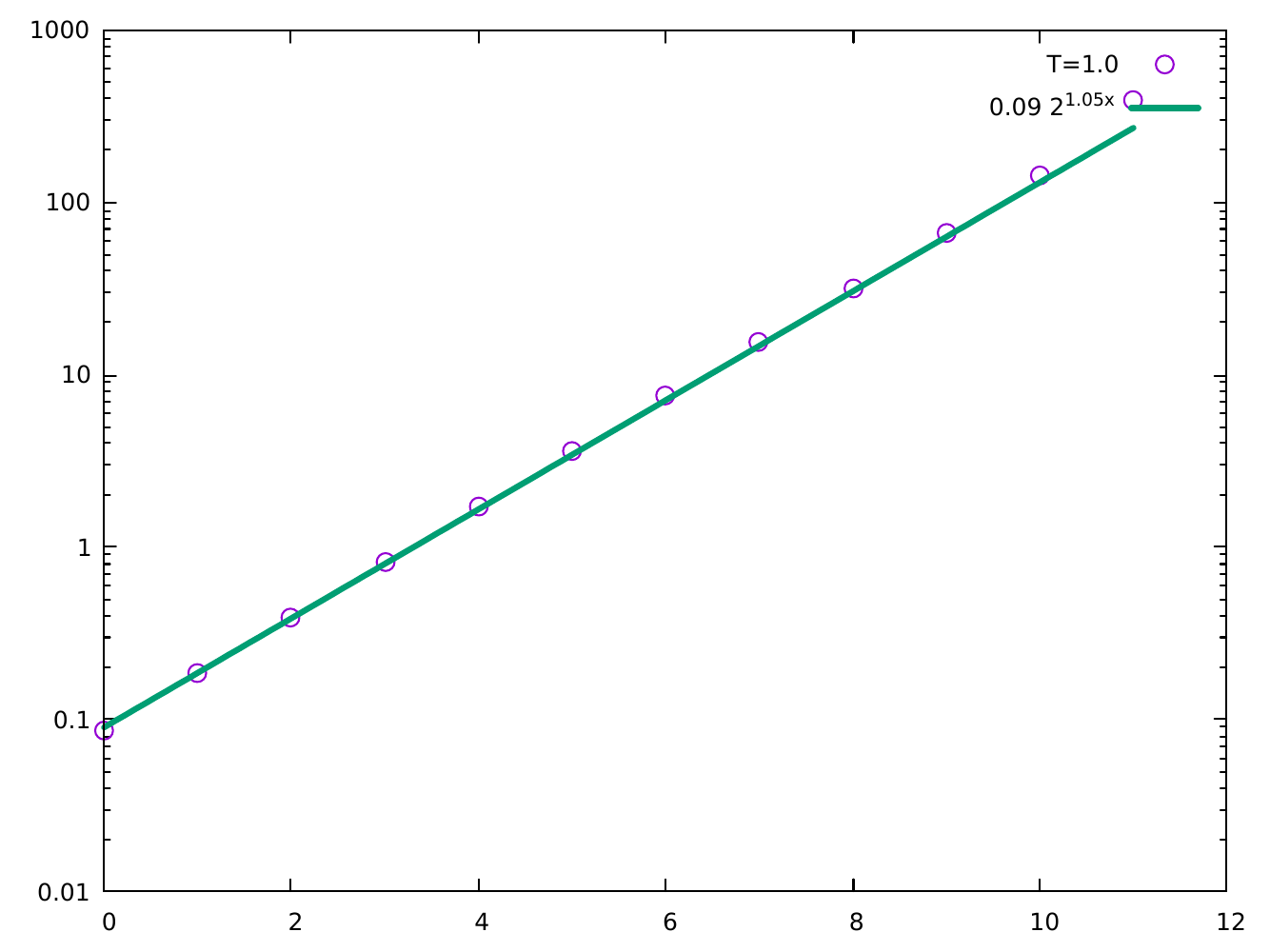}
        \caption{mean-variance to coarse-graining level at $T=1.0$}
        \label{fig:IL-b}
    \end{subfigure}
    \begin{subfigure}[b]{0.3\textwidth}
        \includegraphics[width=\textwidth]{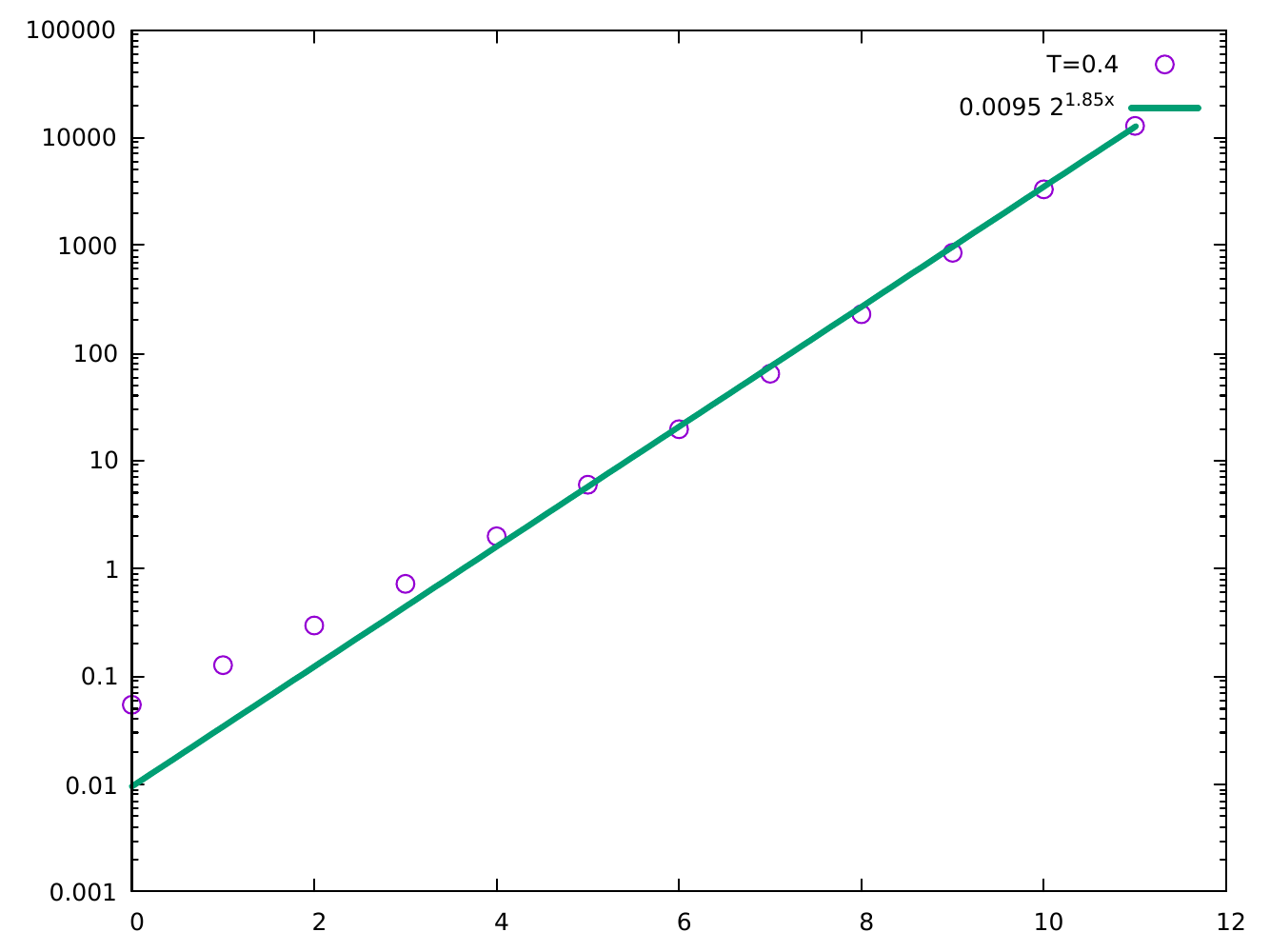}
        \caption{mean-variance to coarse-graining level at $T=0.4$}
        \label{fig:IL-c}
    \end{subfigure}
    \caption{(a) The transient behavior becomes sharper repeating the coarse-graining steps. (b) At $T=1.0$, the expoenent near $1$, while (c) it is meaningfully larger than $1$ at low temperature $T=0.4$. }
\end{figure*}

We construct glass using a Kob-Andersen binary Lennard-Jones mixture model with $80:20$ ratio. The density is fixed as $1.2$ using a Nose-Hoover thermostat. The LJ parameters are are $\sigma_{AA} = 1.0, \sigma_{AB} = 0.8, \sigma_{BB} = 0.88, \epsilon_{AA} = 1.0, \epsilon_{AB} = 1.5$, and $\epsilon_{BB} = 0.5$ in reduced unit.  
$m_A = m_B = 1$. Time is scaled by $\sqrt{(m A \sigma_{AA})^2 / \epsilon_{AA}} = \tau$. 
The cut off of the potential is $2.5 \sigma_{AA}$ and set to zero at cutoff distance.

Figure~\ref{fig:IL-a} shows the CODs at different coarse-graining levels. As the coarse-graining steps are repeated, the intermediate regions are relocated to one of the states. As a result, the transient region becomes sharper.

One virtue of our method is that it provides information about the size of collective behavior or domain regardless of the domain shape. At higher temperatures, we find that the nonlinear behavior in COD disappears at large coarse-graining steps, which justifies the size of the correlation length. 
For example, at temperature $T=0.47$, the nonlinear behavior holds for the first $4$ coarse grain steps but disappears afterward. It implies that the size of the correlated behavior is about $2^4$ particle clusters. 

In Figures~\ref{fig:IL-b} and \ref{fig:IL-c} we estimate the exponents of the power-law below and above the transient temperature.
The exponent is almost $1$ at high temperatures, as expected but is as large as $1.85$ at $T=0.4$. It implies that at low temperatures the correlation is very persistent over the whole system. 

\section{Discussion}
We develop a coarse-graining protocol for extracting collective features of time-series data, including simulation trajectories. 
Using this approach, we identify the existence of the collective behaviors of glass and characterize the size of the correlated particles near the transient temperature. 

Recently, there have been attempts to identify ambiguous collective behaviors in fluids using machine learning technique~\cite{lee2020two,ha2019universality,ha2018widom}. 
Despite those studies being inspiring, the transient behaviors are well justified. Perhaps our approach may improve such studies. 

At first, we start with a nearly Gaussian data set. Total information loss is minimized by maximizing the correlation between a data point and its pair while minimizing the correlations between the pairs and other data points. This approach effectively reduces the dimensionality of the data while preserving the most essential features. We extend the coarse-graining rule from linear analysis to non-linear data analysis, which is eventually expressed as an information bottleneck. 

\bibliography{mybib}

\end{document}